\documentclass{mn2e}
\usepackage{times}
\usepackage{epsf}

\title[Longest superhumps in TV Col]
{A 6.3-h superhump in the cataclysmic variable TV~Columbae: the longest yet
seen}

\author[A. Retter et al.]
{A. Retter$^{1,2}$\thanks{Email: retter@physics.usyd.edu.au; 
ch@astro.keele.ac.uk; tau@not.iac.es; timn@astro.ex.ac.uk; 
bedding@physics.usyd.edu.au; bembrick@ix.net.au; fco@xtra.co.nz},
C. Hellier$^2$, 
T. Augusteijn$^3$, T. Naylor$^{4,2}$, T.R. Bedding$^{1}$, C. Bembrick$^5$, 
\newauthor J. McCormick$^6$ and F. Velthuis$^6$ 
\smallskip 
\\
$^{1}$School of Physics, University of Sydney, 2006, Australia\\
$^{2}$Department of Physics, Keele University, Keele, Staffordshire 
ST5 5BG\\
$^{3}$Nordic Optical Telescope, Apartado 474, 38700 Santa Cruz de La Palma, 
Canary Islands, Spain\\
$^{4}$School of Physics, University of Exeter, Stocker Road, Exeter EX4 4QL\\
$^{5}$Mt Tarana Observatory, PO Box 1537, Bathurst, NSW 2795, Australia\\
$^{6}$Centre for Backyard Astrophysics (Pakuranga), Farm Cove Observatory, 
2/24 Rapallo Place, Farm Cove, Auckland, New Zealand\\
}

\date{Accepted 2002 December 4. Received 2002 November ??; in original form 
2000 February 2}

\begin{document}

\maketitle

\begin{abstract}

We present results from a two week multi-longitude photometric campaign
on TV~Col held in 2001 January. The data confirm the presence of a 
permanent positive superhump found in re-examination of extensive archive 
photometric data of TV~Col. The 6.3-h period is 15 per cent longer than the 
orbital period and obeys the well known relation between superhump period 
excess and binary period. At 5.5-h, TV~Col has an orbital period longer 
than any known superhumping cataclysmic variable and, therefore, a mass 
ratio which might be outside the range at which superhumps can occur 
according to the current theory. We suggest several solutions for this 
problem. 

% TV~Col also had a mini-outburst at the middle of the 2001 January
% run. The 6.3-h period is not seen in the data before the outburst. The
% negative superhump phase was changed through the outburst by about 0.25.

\end{abstract}


\section{Introduction}
\subsection{Permanent superhumps}

Patterson \& Richman (1991) initially suggested the term `permanent
superhump' for the subclass of cataclysmic variables (CVs) having
quasi-periodicities slightly different from their binary orbital periods.
Unlike SU~UMa systems (see Warner 1995 for a review of SU~UMa systems and
CVs in general), which show this behaviour only during superoutbursts,
permanent superhump systems show the phenomenon during their normal
brightness state.  However, their amplitudes are highly variable and are
sometimes below the detection limits, so the term `permanent superhump' is
somewhat misleading.

Whitehurst \& King (1991) suggested that superhumps occur when the
accretion disc extends beyond the 3:1 resonance radius.  According to Osaki
(1996), permanent superhumpers differ from other subclasses of non-magnetic
CVs in having relatively short orbital periods and high mass-transfer
rates, resulting in accretion discs that are thermally stable but tidally
unstable.  Retter \& Naylor (2000) provided observational support for this
idea.

The `positive superhump', a periodicity that is a few per cent larger 
than the orbital period, is explained as the beat period between the 
binary motion and the precession of an eccentric accretion disc in the 
apsidal plane. Periods slightly shorter than the orbital periods have 
also been seen in several systems.  They are known as `negative 
superhumps', and may be generated by the nodal precession of the 
accretion disc (Patterson et al. 1993; Patterson 1999).  However, there 
are some theoretical difficulties with this idea (Murray \& Armitage 1998;
Wood, Montgomery \& Simpson 2000; Murray et al. 2002).  Superhumps were
also associated with the formation of a spiral structure in the accretion
disc (Steeghs, Harlaftis \& Horne 1997; Baba et al. 2002).

Observations of positive superhumps have shown a roughly linear
relationship between the period excess, expressed as a fraction of the
binary period, and the binary period itself (Stolz \& Schoembs 1984).
Negative superhumps seem to obey a similar rule (Patterson 1999).

% The X-ray period has not been clearly seen in any optical data set.

% \note{this para. not needed -- delete?}
% \Mini-outbursts with amplitudes of 1--2 mag have also been detected in the
% \light curve of TV~Col (Szkody \& Mateo 1984; Schwartz et al.\ 1988; Hellier
% \\& Buckley 1993; Augusteijn et al.\ 1994). Recent photometry suggests they
% \are quite frequent (Kato 2001a; 2001b).  Szkody \& Mateo and Schwartz et
% \al.\ explained this behaviour in terms of disc instability, similar to the
% \outbursts observed in dwarf novae. The difference between these
% \short-lived, low-amplitude mini-outbursts and the longer, stronger
% \dwarf-nova outbursts was explained as a result of the truncation of the
% \inner accretion disc, which is believed to occur in intermediate polars due
% \to the strength of the magnetic field of the primary white dwarf. Hellier
% \\& Buckley, however, concluded that during an outburst in TV~Col, the
% \brightness of the bright spot was significantly enhanced. Therefore, they
% \suggested that the mini-outbursts are mass transfer events. Angelini \&
% \Verbunt (1989) applied the thermal disc instability and the mass transfer
% \models to outbursts in intermediate polars, and showed that truncated
% \accretion discs in these systems can indeed cause short outbursts in both
% \models.

% Also shown in Table~1 are the observations in 2001. These were divided into 
% two sets -- before the outburst (five nights), and afterwards (eight nights).

\begin{table*}
\begin{minipage}{300 mm}

\caption{\label{table.all} Photometric observations of TV Col}
\begin{tabular}{@{}lccccccccc@{}}

\hline

Set & Month/year & Nights & Site     & Telescope size  & Detector   & Filter/s  & Exposures & Mean night & Number of \\
    &            &        &           & [m]            &            &           & [s]      & length [h] & outbursts \\
\hline

1   & 12/1985      & 6      & ESO      & 0.91           & photometer & V,B,L,U,W & 16        & 4.2 (gaps) & 0       \\

2   &12/1985--1/1986  & 13     & SA       & 0.75+1.0       & photometer & white     & 2        & 3.0        & 0      \\

3   &11--12/1987    & 16     & ESO      & 0.91           & photometer & V,B,L,U,W & 16        & 5.4 (gaps)  & 3      \\

4   & 12/1987      & 9      & ESO      & 0.91           & photometer & V,B,L,U,W & 16        & 5.0        & 0       \\

5   & 11/1988      & 7      & ESO      & 0.91           & photometer & V,B,L,U,W & 16        & 4.4 (gap)  & 0       \\

6   & 1/1989       & 6      & SA       & 1.0            & CCD        & white     & 40        & 7.3        & 0      \\

7   & 1/1991       & 9      & SA       & 0.75           & photometer & B,R       & 4         & 4.8        & 0      \\

8   & 12/1991      & 11     & SA       & 0.75           & photometer & white     & 10        & 4.0        & 1      \\

9   & 1/2001       & 5      & SA+AU+NZ & 0.75+0.40+0.25 & CCD        & white     & 20-90     & 11.8       & 0      \\

10  & 1/2001       & 8      & SA+AU+NZ  & 0.75+0.40+0.25 & CCD        & white     & 20-90     & 10.6       & 0      \\

%1   & 12/85      & 6      & ESO       & 0.91 & photometer & Walraven (V,B,L,U,W) & 16        & 4.2 (gaps) & 0       \\
%2   &12/85-1/86  & 13     & SA      &0.75+1.0& photometer & white (blue sensitive)& 2        & 3.0        & 0      \\
%3   &11-12/87    & 16     & ESO       & 0.91 & photometer & Walraven (V,B,L,U,W) & 16        & 5.4 (gaps) & 2 or 3  \\
%4   & 12/87      & 9      & ESO       & 0.91 & photometer & Walraven (V,B,L,U,W) & 16        & 5.0        & 0       \\
%5   & 11/88      & 7      & ESO       & 0.91 & photometer & Walraven (V,B,L,U,W) & 16        & 4.4 (gap)  & 0       \\
%6   & 1/89       & 6      & SA        & 1.0  & CCD        & white                & 40        & 7.3        & 0      \\
%7   & 1/91       & 9      & SA        & 0.75 & photometer & B,R (broad bands)    & 4         & 4.8        & 0      \\
%8   & 12/91      & 11     & SA        & 0.75 & photometer & white                & 10        & 4.0        & 1      \\
%9   & 1/01       & 5      & SA+AU+NZ  & 0.75 & CCD        & white                & 20-60     & 11.8       & 0      \\
%10  & 1/01       & 8      & SA+AU+NZ  & 0.75 & CCD        & white                & 20-60     & 10.6       & 0      \\

\hline

\end{tabular}
\end{minipage}
\end{table*}

\subsection{TV~Col}
%\subsection{Periodicities in TV~Col}

% The light curve of TV~Col, a 14th mag CV, shows multiple periodicities 

TV~Col is a 14th mag CV at a distance of about 370 pc (McArthur et al.\
2001). Its light curve shows multiple periodicities that have attracted
many observers. Motch (1981) found periodicities of 5.2\,h and 4\,d from
photometry taken over 10 nights.  A radial velocity study by Hutchings et
al.\ (1981) confirmed these periods and detected another at 5.5 h, which
they identified as the orbital period.  They also pointed out that the 4-d
period is exactly the beat periodicity between the other two periods. The
5.2-h period was interpreted as the spin period of a magnetic white dwarf,
leading to a classification of the object as an intermediate polar (for
reviews of intermediate polars, see Patterson 1994; Warner 1996; Hellier
1996). The detection of a 32-m period in x-ray observations (Schrijver et
al.\ 1985; Schrijver, Brinkman \& van der Woerd 1987) confirmed the
intermediate polar nature of TV~Col, but left the 5.2-h period
unidentified.

Further observations confirmed the presence of the three periods in the
optical regime (Barrett, O'Donoghue \& Warner 1988; Hellier, Mason \&
Mittaz 1991; Hellier 1993; Augusteijn et al.\ 1994). The orbital period was
also detected in the ultraviolet (Bonnet-Bidaud, Motch \& Mouchet 1985). It
was found that the 5.2-h period is not stable (e.g. Augusteijn et
al.\ 1994), suggesting that it is a superhump period. The fact that it is
shorter than the binary period would make it a negative superhump.  Barrett
et al.\ (1988), however, discussed other possible models for the system.

\subsection{The significance of superhumps in TV~Col}

Superhumps have only been observed in CVs with orbital periods below about
3.7\,h.  Since the orbital period of TV~Col is 5.5 h, the interpretation of
its 5.2-h period as a negative superhump would make this system an extreme
case.  There is known to be a strong correlation in CVs between the orbital
period and the mass of the secondary star (Smith \& Dhillon 1998), which
arises because systems with longer orbital periods have larger separations
and thus require more massive secondaries to fill their Roche-lobes.  The
5.5-h orbital period of TV~Col implies a secondary star of $\sim 0.6
M_{\odot}$ -- much more massive than in other superhumping systems. A
typical white dwarf in CVs has a mass of $\sim 1.0 M_{\odot}$, which means
the mass ratio, $q=M_{\rm donor}/M_{\rm compact}$, in TV~Col could be about
0.6 or even higher.  This value significantly exceeds the theoretical limit
for superhumps -- $q\la$0.33 (Whitehurst 1988; Whitehurst \& King 1991;
Murray 2000). TV~Col thus offers a unique opportunity to test the
predictions of the models for precessing accretion discs.

There seems to be a strong connection between positive and negative 
superhumps. Light curves of many permanent superhumpers show both types 
of superhumps (Patterson 1999; Arenas et al.\ 2000; Retter et al.\ 2002). 
In addition, period deficits in negative superhumps are about half the 
period excesses in positive superhumps (Patterson 1999): 
$\epsilon_{\rm negative}\approx -0.5 \epsilon_{\rm positive}$, where 
$\epsilon =(P_{\rm superhump} - P_{\rm orbital})/P_{\rm orbital}$. 

This prompted us to examine available photometry of TV~Col for positive
superhumps, which would be predicted to have a period near 6.3\,h.  We
indeed found such evidence (Retter \& Hellier 2000; Retter et al.\ 2001),
which is presented here.  At the request of a referee, we then obtained new
multi-site photometry on TV~Col in 2001 January, which confirmed the 6.3-h
period.

\section{Photometric data}

\subsection{Published observations}

We have re-analysed existing optical photometry that was presented by
Barrett et al.\ (1988); Hellier et al.\ (1991); Hellier (1993); Hellier \&
Buckley (1993) and Augusteijn et al.\ (1994).  This data set contains 77
nights of observations obtained during 1985--1991 over eight separate runs,
using the 0.75-m and 1-m telescopes at the South African Astronomical
Observatory (SAAO) and the Dutch telescope at the European Southern
Observatory (ESO).  The main properties of these data are summarized in the
first eight rows of Table~\ref{table.all}.

\subsection{New observations}

In 2001 January 2--15 we obtained optical photometry of TV~Col from three
locations: Sutherland, South Africa (SA) using the 0.75-m telescope with
the UCT CCD photometer; Mt Tarana Observatory, Australia (AU; 0.40-m
telescope; ST8 CCD) and Auckland, New Zealand (NZ; 0.25-m telescope; ST6
CCD).  No filters were used and typical exposure times were a few tens of
seconds.  Altogether, more than 150 hours of useful data were collected.

For the SA data, PSF-fitting photometry was carried out using a comparison
star $9''$ NW of TV~Col.  Aperture photometry yielded similar results, with
differences typically less than 0.02\,mag.  The data from AU and NZ were
reduced using aperture photometry, with GSC7059:486 (AU) and GSC7059:754
(NZ) as reference stars.  Since different comparison stars were used at the
three sites, a linear fit was subtracted from each whole data set before
combining them. 

%\note{Was this done night by night, or to the whole sets?}

Table~\ref{table.new} shows the log of the observations, while Figs.~1
and~2 show the resulting light curve.  We see that an outburst of
$\sim$1\,mag occurred on 2001 January 7, which lasted between 7.6 and
30\,h.  These data were excluded from the analysis, and the remainder were
divided into two sets: before the outburst (Set~9), and afterwards
(Set~10).  The properties of these sets are included as the last two rows
in Table~\ref{table.all}.

\begin{figure*}

\centerline{\epsfxsize=4.0in\epsfbox{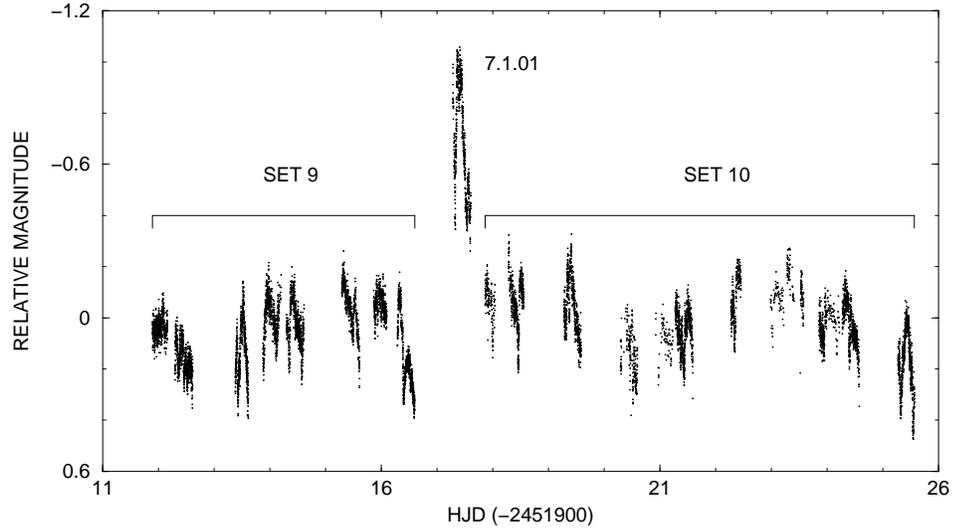}}

\caption{Light curve of the 2001 January observations (`clear filter').
A mini-outburst occurred at day 6 of the run.}

\end{figure*}

\begin{table}
\caption{\label{table.new} New photometric observations}
%\caption{The observations time table}
\begin{tabular}{@{}lccccc@{}}

UT     & Time of Start &Run Time&Points&Site & Notes    \\
Date   &    [HJD]      & [h]    &number&     &          \\
\\

020101 & 2451911.895   &  6.5   &  541  & NZ &          \\
020101 & 2451911.963   &  3.5   &  30   & AU &          \\
020101 & 2451912.298   &  7.6   &  796  & SA &          \\
030101 & 2451913.382   &  5.6   &  612  & SA &          \\
040101 & 2451913.879   &  6.9   &  557  & NZ &          \\
040101 & 2451913.930   &  6.6   &  77   & AU &          \\
040101 & 2451914.299   &  7.6   &  680  & SA &          \\
050101 & 2451915.292   &  7.7   &  583  & SA &          \\
060101 & 2451915.869   &  5.5   &  456  & NZ &          \\
060101 & 2451916.291   &  7.5   &  774  & SA &          \\

070101 & 2451917.287   &  7.6   &  650  & SA & outburst \\

080101 & 2451917.864   &  4.2   &  150  & NZ &          \\
080101 & 2451918.286   &  6.5   &  558  & SA &          \\
090101 & 2451919.281   &  7.4   &  630  & SA &          \\
100101 & 2451920.296   &  7.2   &  225  & SA &          \\
110101 & 2451920.924   &  7.4   &  69   & AU &          \\
110101 & 2451921.278   &  7.6   &  726  & SA &          \\
120101 & 2451922.278   &  4.3   &  308  & SA &          \\
130101 & 2451922.986   &  5.7   &  65   & AU &          \\
130101 & 2451923.287   &  7.0   &  136  & SA &          \\
140101 & 2451923.860   &  3.2   &  235  & NZ &          \\
140101 & 2451923.986   &  7.3   &  85   & AU &          \\
140101 & 2451924.277   &  7.2   &  678  & SA &          \\
150101 & 2451925.276   &  7.1   &  756  & SA &          \\

\end{tabular}
\end{table}

\begin{figure}

\centerline{\epsfxsize=3.0in\epsfbox{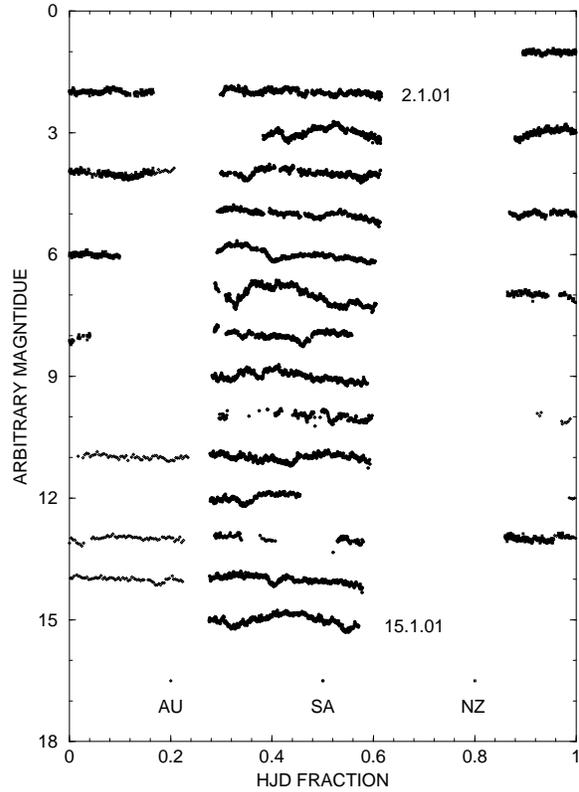}}

\caption{Light curve of the 2001 January observations (`clear filter'). 
Single nights from the three sites are plotted vertically. An offset of 
1 mag was imposed between successive nights}

\end{figure}

\section{Analysis}

\subsection{General remarks}

When searching for a 6-h superhump period in the light curve of a variable
star there are several complications.  Firstly, CVs tend to have
night-to-night secular variations of a few tenths of a magnitude that are
not related to the periodicity (e.g., Retter, Leibowitz \& Naylor 1999).
Therefore, very often the best detrending of the data for periodogram
analysis is the subtraction of the nightly means / trends. Removing the
mean or trends eliminates noise in the power spectrum from this effect.
Obviously, observations shorter than the period might be badly normalized.
Secondly, superhump periods are not stable.  Thirdly, the mini-outbursts
observed in TV~Col (Section~2.2 and Fig.~1; Szkody \& Mateo 1984; Schwartz 
et al.\ 1988; Hellier \& Buckley 1993; Augusteijn et al.\ 1994) pose 
further difficulties because they can alter the phase of the superhump 
period (Hellier \& Buckley 1993).  Therefore, the data are best analyzed in 
subsets spanning short intervals ($\la$2 weeks).

%($\la$2--3 weeks)

% (Szkody \& Mateo 1984; Schwartz et al.\ 1988; Hellier \& Buckley 1993; 
% Augusteijn et al.\ 1994)

Most subsets shown in Table~\ref{table.all} are not ideal for the search
for a 6 h periodicity as the observations were typically shorter than one
cycle, had long gaps during the nights, extended over a long interval of
time, or had outbursts.  The best sets for our purpose are the two new ones
(9 and 10) and Set~6, as these all consist of long successive nights.  They
also have the advantage of having been carried out with CCDs (rather than
photometers).

As we have discussed previously, Sets~1--8 contain evidence for a
periodicity of 6.3\,h, which we interpreted as being a positive superhump
(Retter \& Hellier 2000; Retter et al.\ 2001).  In the new data, this
periodicity is strongly seen in Set~10.

% no outburst occurred during the runs

\subsection{Set~10}

% The new periodicity is the most significant in Set~10. In 

In Fig.~3a we present the power spectrum (Scargle 1982) of the data in 
Set~10.  Note that no de-trending method was necessary. In addition to the 
three known optical periods (5.5 h, 5.2 h and 4 d, marked as f$_{1}$, f$_{2}$ 
and f$_{3}$) and their 1 d$^{-1}$ aliases, there is a fourth peak (labelled 
f$_{4}$) and its 1 d$^{-1}$ aliases.  This peak is stronger than f$_{1}$, 
and is the fourth highest peak in the graph after f$_{2}$, f$_{3}$ and 
a 1 d$^{-1}$ alias of f$_{2}$.  After fitting and subtracting the three 
known frequencies (f$_{1}$, f$_{2}$ and f$_{3}$) and detrending by 
subtracting a linear term from each night, f$_{4}$ becomes the strongest 
peak in the residual power spectrum (Fig.~3b).

% Note that no de-trending method was necessary and the result after
% subtracting the mean or the linear trend from single nights is similar.
% \note{unclear}

\begin{figure}

\centerline{\epsfxsize=3.0in\epsfbox{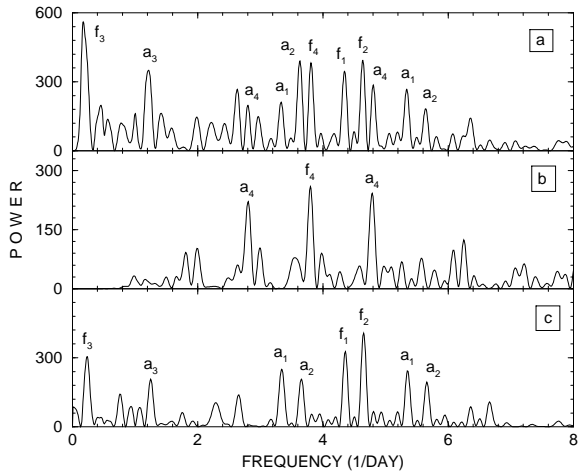}}

\caption{Power spectra of Set~10. (a) Raw data; In addition to the 
three previously known periods -- the orbital period (marked as f$_{1}$), 
the negative superhump (f$_{2}$) and the beat between the two (f$_{3}$), 
there is a fourth structure of peaks centered around 6.3~h (f$_{4}$); 
`a$_{i}$' ($i=1-4$) represent 1 d$^{-1}$ aliases of `f$_{i}$' correspondingly;
(b) After fitting and subtracting f$_{1}$, f$_{2}$, f$_{3}$ and the 
nightly trends, the period is still present, and becomes the strongest peak 
in the graph.
(c) A synthetic light curve, consisting of sinusoids of the three 
previously known optical periods (5.2 h, 5.5 h, 4 d) (plus noise) sampled 
as the data thus illustrating the window function. This test shows that 
aliases of the three known periods cannot explain the f$_{4}$ peak.}

% planted in the window function plus noise. No significant additional 
% structure / periodicity is seen.}

\end{figure}

In Fig.~4 we show the data of Set~10 folded on the 6.3-h period after the
three previously known periods (5.2 h, 5.5 h and 4-d) were removed. The
peak-to-peak amplitude of the sinusoidal fit to the light curve is
$0.08\pm0.01$ mag.

% (2.5$\sigma$) mag. 

\begin{figure}

\centerline{\epsfxsize=3.0in\epsfbox{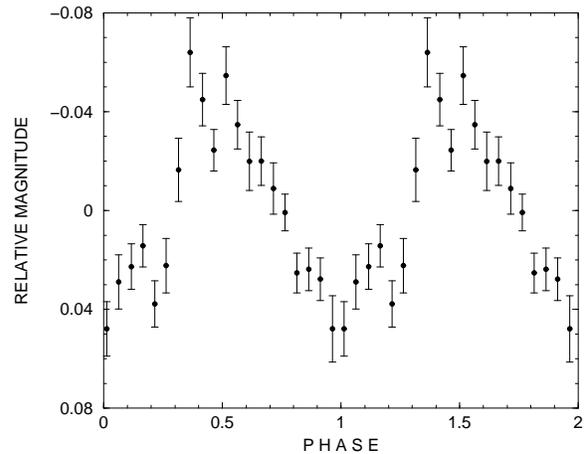}}

\caption{The 2001 January data (after the outburst) folded on the 6.3-h 
period and binned into 20 bins. Two cycles are shown for clarity. The 
bars represent 1-$\sigma$ errors in each bin.}

\end{figure}

\subsection{Other subsets}

Figure~5 presents power spectra of the five best sets among all ten. Sets 
1, 2, 3, 5 and 8 were rejected because either outbursts occurred during the 
observations (Sets~3 and 8), and the data between outbursts were too scarce; 
the mean night length was too short (Sets~1, 2 and 8) or there were long gaps 
during the runs (Sets~1, 3 and 5).  The three previously known periods (5.2 h, 
5.5 h, 4 d) have been removed, as have the nightly trends.  In the residual 
power spectra of Sets~4, 6 and 7, the highest peak (or a 1 d$^{-1}$ alias) 
is compatible with the peak from Set~10. It is absent from Set 9, which is 
one of the best data sets, but only gives an upper limit of 0.02 mag on the 
peak-to-peak amplitude of the 6.3-h period in these data. We return to this 
issue below.

% Note, however, that in Sets~2 and 5, the mean length of the nightly 
% observation was shorter than one cycle of the periodicity.  

% Set~9 is one of the best data sets, but only gives an upper limit of 0.02 
% mag on the full amplitude of the 6.3-h period in these data. We return to 
% this issue below.

% This issue is further discussed below.

\begin{figure}

\centerline{\epsfxsize=3.0in\epsfbox{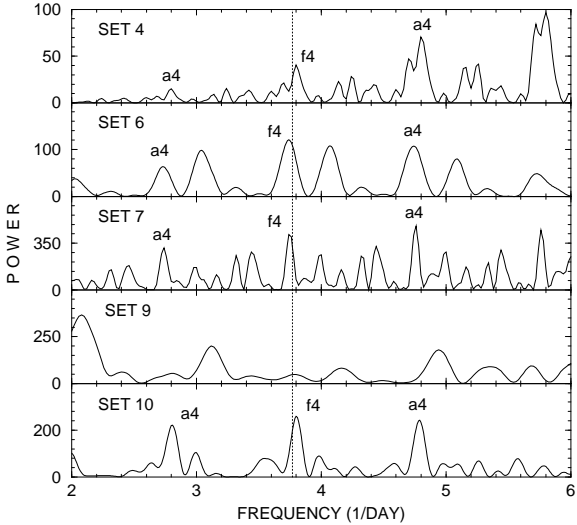}}

\caption{Power spectra of the best five subsets after the removal of the 
three previously known periods (5.2 h, 5.5 h, 4 d) and a linear fit from 
each night. In Sets 4, 6 and 7 the strongest peak (or a 1 d$^{-1}$ alias) 
is consistent with the location of the peak found in Set~10 (which is 
identical with Fig.~3b). Its 1 d$^{-1}$ aliases (a$_{4}$) are also marked. 
Set~9 do not show this pattern despite presumably being one of the best 
sets. This issue is further discussed below. The vertical dotted line 
shows the mean of the period found from the four other data sets.}

% without apparent outbursts nor gaps during the runs,

% The other three data sets do not show this pattern, however, two of them 
% (2 and 5) are of low quality. The vertical dotted line shows the 
% mean of the period found from the five data sets.} 

%See text for more details.}

\end{figure}

In Sets 4, 6, 7 and 10 the f$_{4}$-peak lies in the range 3.74--3.84
d$^{-1}$ ($0.2639\pm0.0035$ d).  The large interval originates from many
observations covering less than a full cycle of the period (thus being
affected by the detrending method used), but might also reflect true
long-term changes in the periodicity.  Note that superhumps in other
systems have shown large changes (e.g., up to more than one per cent in
V603~Aql -- Patterson et al.\ 1997).

% The vertical dotted line in Fig.~5 shows the mean of these values.

\subsection{Tests}

To reject the possibility that individual nights may be responsible for the
appearance of the period, we calculated power spectra for the data from
Sets~6 and 10 while rejecting one night at a time.  The suspected period
survived this test and was one of the four highest peaks, together with the
5.2-h, 5.5-h and 4-d periods (besides 1 d$^{-1}$ aliases). In addition,
when the data in Set~10 were divided into two independent sets consisting
the even nights and odd nights, the 6.3-h period appeared in both sets.

We investigated other possible spurious sources of the candidate period, in
a series of tests detailed below.  We first assessed whether the period
could be an artefact of the window function in combination with the known
periods. We then tested whether noise, either uncorrelated or correlated,
could be the source of the periodicity.  We then considered the possibility
that extinction may be responsible for the appearance of the period.
Finally, we comment on the fact that the superhump periodicity in TV Col
was discovered in published observations and subsequently confirmed by the
new data.

\subsubsection{The window function}

Could the new peak in the power spectrum be an artefact of the window
function?  To check this we created a noiseless simulation of two of the
best sets (6 and 10).  A synthetic light curve was built using three
sinusoids at the orbital period, the negative superhump period and at 
the beat period between the two.  These sinusoids were given the same 
amplitudes they have in the data, and sampled according to the window
function.  There was no evidence for significant power at the proposed
period.

\subsubsection{Uncorrelated noise}

To check whether uncorrelated noise could be responsible for the presence
of the candidate period, we added noise to the model light curves of Sets~6
and 10 (previous section).  The noise in the original data was defined as
the root mean square of the data minus the three periods modelled. We then
searched for the highest peak in a small interval (3.72--4.00\,d$^{-1}$)
around the candidate period.  In 1000 simulations, no peak reached the
height of the candidate periodicity.  An individual example of a simulation
for Set~10 is shown in Fig.~3c. In a further test we mimicked the technique
shown in Fig.~3b, by subtracting the previously known periods (5.2 h, 5.5
h, 4 d) from our model light curves, after imposing an error of 0.005
d$^{-1}$ in the periods. We then checked the height of the highest peak in
the resulting power spectrum. In 1000 simulations, no peak reached the
level of the candidate periodicity.

\subsubsection{Correlated noise}

We also tried to assess the probability that correlated noise could be
responsible for the candidate periodicity. In the absence of a model for
the correlated noise, the best test is to use the repeatability between
different data sets. Given that we found a period in Set~10, and assuming
that the nearby peak in Set~4 represents the same period, we can ask how
likely it is that the strongest period in the other data sets (after the
previously known periods had been subtracted) would be consistent with it.
The probability of the highest peak in another data set being, by chance,
compatible with the candidate period in Set~10 is 0.1. This was calculated
from (i) the frequency range for the candidate period on the assumption
that the peaks in Set~4 and Set~10 set this range, which implies that the
peaks are compatible if they are within 0.1 d$^{-1}$, and (ii) the range
over which it could occur taken as the spacing of the 1 d$^{-1}$ aliases (1
d$^{-1}$ is the maximum range over which periods are truly independent).
The period discovered in Set~10 was seen in two of the remaining three data
sets (besides Set~4). We thus used the binomial distribution to find that
the probability of this occurring by chance was only about 2.8 per cent.

% We also tried to assess the probability that correlated noise could be
% responsible for the candidate periodicity. In the absence of a model for
% the correlated noise, the best test is to use the repeatability between
% different data sets. Given that we found a period in Set~10, and assuming
% that the nearby peak in Set~1 represents the same period, we can ask how
% likely it is that the strongest period in the other data sets (after the
% previously known periods had been subtracted) would be consistent with it.
% The probability of the highest peak in another data set being, by chance,
% compatible with the candidate period in Set~10 is 0.1. This was calculated
% from (i) the frequency range for the candidate period on the assumption
% that the peaks in Set~1 and Set~10 set this range, which implies that the
% peaks are compatible if they are within 0.1 d$^{-1}$, and (ii) the range
% over which it could occur taken as the spacing of the 1 d$^{-1}$ aliases (1
% d$^{-1}$ is the maximum range over which periods are truly independent).
% The period discovered in Set~10 was seen in three of the remaining six data
% sets (besides Set~1). We thus used the binomial distribution to find that
% the probability of this occurring by chance was only about 0.1 per cent.

% However, two of the sets that do not show the period (1 and 2) are those 
% with the shortest runs each night (Table~1), and thus the lowest data 
% quality. Thus the 99.8 per cent significance value should be regarded as 
% a lower limit.

\subsubsection{Extinction effects}

The observations were made in white light, and the differential photometry
was performed relative to three comparison stars that are almost certainly
redder than TV~Col.  Since there are very few blue stars available in a
typical CV field, this is a standard procedure in CCD photometry of CVs.
The colour difference may result in extinction variations with airmass on
the data.  For a single observing site this could introduce apparent
periodicities in the data which should approximately be equal to a fraction
of a day, i.e. a period of 12 h, 8 h, 6 h, etc. This might explain the
detection of a 6.3 h, although the typical length of the observing runs
would more favour a period of 8 or 12 h.  However, it is very unlikely 
that two different data sets would be affected by extinction variations 
such that the same period is found in both.  Furthermore, the observations
in data set 10 were in fact obtained at different observing sites, where
the source will cross the meridian at very different times, which are far
from being equal an integral number of the new period we found.

It is important to realise that variations in the data due to extinction
are not intrinsic to the source, i.e. this will distort the light curve 
and will actually affect the detection of any period that is present in 
the source.  In fact, we do detect the previously known 5.2 and 5.5-h 
periods at the correct positions in the power spectra, so this effect 
cannot be strong.  As shown in Fig.~3, we detect the 5.2, 5.5 and 6.3-h 
periods at very similar strengths in data set 10, and it is unclear how 
one could reject the detection of the 6.3-h period without having to 
discard the 5.2 and 5.5-h periods as well.  Nevertheless, as a final 
test we fitted and subtracted a sinusoid corresponding to the sidereal 
day ($23^{h}56^{m}4^{s}$) from the SA data and checked the resulting 
power spectrum.  The candidate period survived this test, which confirmed 
that extinction effects are not responsible for the presence of the 6.3-h 
period.

\subsubsection{Confirmation from new observations}

Finally, we note that we have previously argued for a 6.3-h periodicity
when we had only the data sets 1--8 (Retter \& Hellier 2000; Retter et al.\
2001).  Its presence has now been confirmed by the new data.  Table~3
summarizes all periods detected in the light curve of TV~Col.

\begin{table}
\caption{The periods of TV~Col}
\begin{tabular}{@{}clll@{}}

\hline

Number & Period [d]     & Nature     & Reference                 \\

\hline

1      & 0.022112(58)   & spin       & Schrijver et al. (1985; 1987) \\
%       &                &            & 1987)                     \\
       &                &            &                           \\
2      & 0.21627(7)     & negative   & Motch (1981)              \\
       & 0.21631(1)     & superhump  & Hutchings et al.\ (1981)  \\
       & 0.216325(1)    &            & Barrett et al.\ (1988)    \\
       & 0.2162774(14)  &            & Hellier et al.\ (1991)    \\
       & 0.2162783(12)  &            & Hellier (1993)            \\
       & 0.216036(93)   &            & Augusteijn et al.\ (1994) \\
       &                &            &                           \\
3      & 0.228600(5)    & orbital    & Hutchings et al.\ (1981)  \\
       & 0.228685(3)    &            & Barrett et al.\ (1988)    \\
       & 0.2285529(2)   &            & Hellier et al.\ (1991)    \\
       & 0.2286034(16)  &            & Hellier (1993)            \\
       & 0.22859884(77) &            & Augusteijn et al.\ (1994) \\
       &                &            &                           \\
4      & 0.2639(35)     & positive   & This work                 \\
       &                & superhump  &                           \\
       &                &            &                           \\
5      & 3.90(15)       & nodal      & Motch (1981)              \\
       & 4.024(4)       & precession & Hutchings et al.\ (1981)  \\
       & 4.02603(3)     &            & Barrett et al.\ (1988)    \\
       & 4.0283(5)      &            & Hellier et al.\ (1991)    \\
       & 3.934(70)      &            & Augusteijn et al.\ (1994) \\

\hline

\end{tabular}

\end{table}

\section{Discussion}

\subsection{The new period}

The photometric data show a periodicity of 0.2639\,d, in addition to 
the previously known periods.  The repeatability of the peak in several
independent data sets makes it highly significant, and it is also
comforting to note that the new period is generally seen best in the 
best data sets.

The 6.3-h period has almost exactly the value predicted from the relation
of Stolz \& Schoembs (1984; updated by Patterson 1999), which is shown in
Fig.~6.  TV~Col has already been classified as a permanent superhump system
because its 5.2-h period was interpreted as a negative superhump
(Section~1.2).  Moreover, the new period and the negative superhump obey
the relation between the two types of superhumps mentioned in Section~1.3.
Further support for this idea (although not completely independently) comes
from the fact that TV~Col obeys a recently-proposed relation between the
orbital period and the ratio between the positive superhump excess over the
negative superhump deficit (Retter et al.\ 2002).  Therefore, the new
period is naturally interpreted as a positive superhump.

\begin{figure}

\centerline{\epsfxsize=3.0in\epsfbox{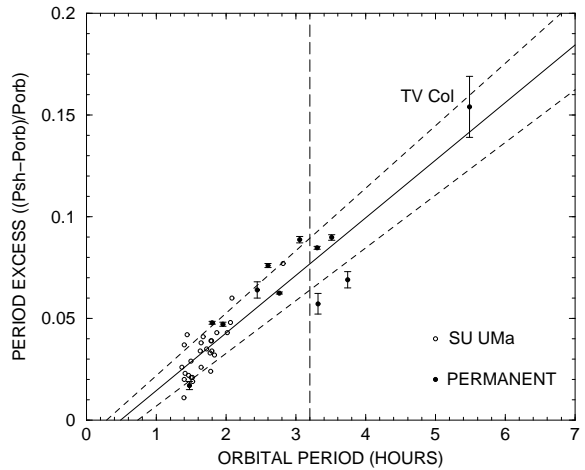}}

\caption{The relation between superhump period excess (over the orbital
period) and orbital period in superhump systems. Empty circles correspond 
to periods in the SU~UMa systems. Filled circles represent the values in 
permanent superhumpers. The data were taken from Patterson (1998; 1999). 
The solid line represents the linear fit to the data. The two tilted dashed 
lines show the 1-$\sigma$ error. There are a few permanent superhump systems 
with periods above the upper edge of the period gap (as defined by Diaz \& 
Bruch 1997), marked by the vertical long-dashed line. The new period 
detected in TV~Col obeys the relation and significantly extends the range of 
superhump periods in CVs.}

\end{figure}

Our result supports the observational connection between positive and
negative superhumps.  Since almost every permanent superhump system that
has been well studied over a few years shows both types of superhumps, we
may speculate that this is the general behaviour among permanent superhump
systems. The idea that the two types of superhumps have a similar physical
origin, namely a precessing accretion disc, is also supported.

\subsection{Changes through the outburst}

TV~Col had a mini-outburst on 2001, January 7 (Section 2.2; Fig.~1). Such 
outbursts are very common in the light curve of TV~Col, but their nature 
is still unclear (Szkody \& Mateo 1984; Schwartz et al.\ 1988; Angelini \&
Verbunt 1989; Hellier \& Buckley 1993; Augusteijn et al.\ 1994; Kato 2001a; 
2001b).

Examination of the data presented above shows that the 6.3-h period appeared 
only in the data taken after the 2001 outburst (Set~10). Before the outburst 
(Set~9) an upper limit of 0.02 mag is found for the peak-to-peak amplitude 
of this periodicity. It is thus possible that the amplitude was increased 
by the outburst by at least a factor of four. Alternatively, the outburst 
triggered the appearance of the positive superhump period. We note that 
amplitudes of permanent superhumps can significantly vary in a single source 
and sometimes the superhumps are even not detected (e.g. Patterson et al.\ 
1997). The amplitude of the negative superhump, on the other hand, was 
slightly lower after the outburst.  Table~4 presents the development of the 
amplitudes of the four periods through the outburst. The data are consistent 
with there being no change in the amplitudes of the 5.5-h and 4-d periods.

% We can also speculate that a second outburst, which can affect the size of 
% the disc and the superhump period, may have occurred during the observations 
% of Set~9.

% TV~Col had a mini-outburst on 2001, January 7 (Section 2.1; Fig.~1).
% Examination of the data presented above shows that the 6.3-h period
% appeared only in the data taken after the outburst (Set~10). Before the
% outburst (Set~9) an upper limit of 0.02 mag is found for the full amplitude
% of this periodicity. The amplitude of the negative superhump, on the other
% hand, was slightly lower after the outburst.  Table~4 presents the
% development of the amplitudes of the four periods through the outburst.
% The data are consistent with there being no change in the amplitudes of the
% 5.5-h and 4-d periods.

\begin{table}
\caption{Changes in the amplitudes of the periods by the outburst}
\begin{tabular}{@{}lcc@{}}

\hline

Period & Peak-to-peak amplitude    & Peak-to-peak amplitude  \\
       & before outburst (Set~9)   & after outburst (Set~10) \\

\hline

5.2 h  & 0.09(1)                   & 0.06(1)                 \\
5.5 h  & 0.10(2)                   & 0.10(2)                 \\
6.3 h  & $<$0.02                   & 0.08(1)                 \\
4 d    & 0.22(6)                   & 0.21(4)                 \\

\hline

\end{tabular}

\end{table}

% positive superhump (6.3 h) & $<$0.02         & 0.08(1)         \\
% negative superhump (5.2 h) & 0.09(1)         & 0.06(1)         \\
% orbital period (5.5 h)     & 0.10(2)         & 0.10(2)         \\
% beat period (4 d)          & 0.22(6)         & 0.21(4)         \\
%                            &                 &                 \\

Previous studies around an outburst in 1991 December showed that the phase
of the 5.2-h period (the negative superhump) was changed by about 0.4 
(Hellier \& Buckley 1993).  Our 2001 data were similarly checked. First, the
ephemeris fitted to the two parts of the data (before and after outburst)
suggested a shift of $\sim$1.4 h (0.27 cycles). To confirm this result,
minima of the negative superhump were calculated from the data after the
subtraction of the other periods (4 d and 5.5 h from both sets, and 6.3 h
only from the latter data).  The O$-$C diagram (Fig.~7) clearly shows a
shift of $\sim$1.2 h or 0.24 cycles.  The data are consistent with no shift
in the phase of the orbital and beat periods.  Note that the errors on the
4-d period and its phase are large as less than two cycles of this
frequency were observed in each set.

\begin{figure}

\centerline{\epsfxsize=3.0in\epsfbox{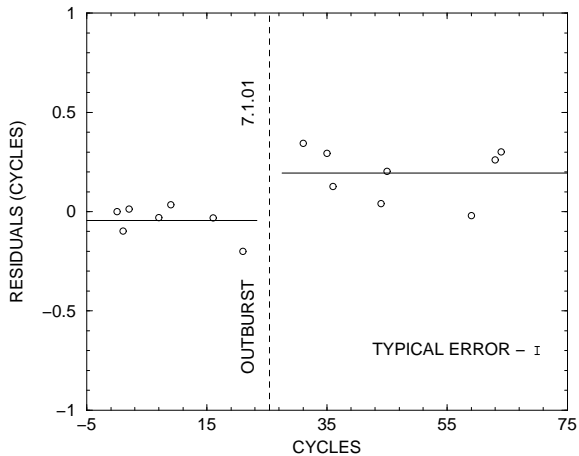}}

\caption{O$-$C diagram of the minima of the negative superhump period. The 
other periods were first subtracted from the data. A parabola was fitted 
to the light curve near the minima to find the points of extrema. The 
dashed vertical line presents the time of the outburst. The two 
horizontal lines show the mean of the points before and after the outburst. 
The phase of the negative superhump was shifted by $\sim$0.25 by the 
outburst.}

\end{figure}

% \subsection{The effect of the outburst}

% During 2001 January, TV~Col had a short-lived mini-outburst (Fig.~1; 
% Sections 2.2, 3.5). The new periodicity appeared in the data following 
% outburst (Set~10), and seems to be absent before the outburst (Set~9). It 
% is possible, however, that the amplitude was below the detection limits at
% that time and was increased by the outburst by at least a factor of four.  
% Alternatively, the outburst triggered the appearance of the positive 
% superhump period. We note that amplitudes of permanent superhumps can 
% significantly vary in a single source and sometimes the superhumps are even 
% not detected (e.g. Patterson et al.\ 1997). We can also speculate that
% a second outburst, which can affect the size of the disc and the superhump 
% period, may have occurred during the observations of Set~9.

% \note{but it did occur!}

% Such outbursts are very common in the light curve of TV~Col
% (Section~1.2).

% The outburst also changed the phase and the 
% amplitude of the negative superhump (Section~3.5).

\subsection{The mass ratio problem}

According to theory, superhumps can only appear in binaries with tidally
unstable accretion discs.  Simulations show that the tidal instability can
only occur if the disc radius exceeds a certain value, the 3:1 resonance
radius.  This implies that eccentric discs (which generate the superhumps)
can be present only in CVs with small mass ratios --
$q$=$M_{donor}/M_{compact}$$\la$0.33 (Whitehurst 1988; Whitehurst \& King
1991; Murray 2000).  Hellier (1993) concluded, however, that the mass ratio
in TV~Col is $q$=0.62-0.93 from a spectroscopic analysis of the system, but
this depended on an interpretation of the radial velocity variation of the
emission lines that may not be correct.

We can estimate the mass ratio in TV~Col by a different method.  The
relation between the orbital period and the superhump period excess
(Fig.~6) suggests that there is a connection between the latter and some
physical quantity. Indeed, this relation was initially explained by Osaki
(1985), who investigated the motion of free particles in the binary
potential.  He considered the axisymmetric part of the tidal perturbation
potential of the secondary star and derived an expression for the
precession rate of the eccentric disc based on a nonresonant free particle
orbit at the disc edge in the first order.  His formula has been commonly
used to obtain the binary mass ratio from the superhump period excess
(e.g. Retter, Leibowitz \& Ofek 1997; Patterson 1998).  The superhump period 
in TV~Col is $15.4\pm1.5$ per cent larger than the orbital period.  Using 
equation (15) of Osaki (1996), we find: $q=0.92\pm0.12$ -- well above the 
0.33 limit suggested by the hydrodynamic simulations, and consistent with 
the values estimated by Hellier (1993) presented above.

% that further assumes that the accretion disc radius equals the critical 
% tidal value ($R_{d}$=0.46a; a is the binary separation),

The mass ratio in TV~Col might thus be inconsistent with the models.  In
the following, we discuss a few possible solutions for this dilemma and
suggest how the observations of TV~Col can be reconciled with the theory.

\subsubsection{TV~Col may be an extreme system}

TV~Col has a binary period of 5.5 h. According to the relation between 
the orbital period and the secondary mass in CVs (Smith \& Dhillon 1998), 
the companion mass is $M_{2}=0.57\pm0.11M_{\odot}$. If the mass of the
primary white dwarf in TV~Col is $M_{1}\approx1M_{\odot}$ (Ramsay 2000), 
the mass ratio is around 0.6 -- still above the theoretical limit. TV~Col 
can have a mass ratio below the critical value only for the secondary mass 
at the bottom of the above range and for a very massive white dwarf near 
the Chandrasekhar mass (1.44$M_{\odot}$). The mass ratio is then 0.32. 
However, for a CV with the limiting theoretical mass ratio of 0.33, a 
superhump period excess of only about 7 per cent is expected according to 
Osaki's equation discussed in the previous section and this is inconsistent 
with the observational value of TV~Col. This fact might indicate that Osaki's 
formula cannot be applied to TV~Col.

%\subsubsection{Using new models}
%%\subsubsection{Using advanced hydrodynamic models}

The equation developed by Osaki is based on a dynamical calculation of
non-resonant particles. Furthermore, it does not assume that the disc
radius exceeds the 3:1 resonance radius. More sophisticated models have
been introduced since. Murray (2000) compared the mass ratios reliably
measured for three eclipsing SU~UMa systems with estimates using the
superhump excess in Osaki's equation and found some inconsistency. He
argued that the use of a gaseous disc (rather than isolated test particles)
modifies the equations. The eccentricity is excited at the 3:1 resonance
and then propagates inwards through the disc. According to his ideas,
pressure forces are an important factor in the calculations, yielding mass
ratios lower than the previous estimates.  Murray also argued that, for
systems with mass ratios exceeding 0.25, the dynamical equations cannot be
applied.  From his simulations (see his fig.~2), and using the observed
superhump period excess of TV~Col, a mass ratio of about 0.3 can be
deduced.  For a white dwarf of $M_{1}=1M_{\odot}$, this implies an
undermassive secondary with $M_{2}=0.3 M_{\odot}$, which would indicate an
evolved star close or near the end of hydrogen burning (see discussions in
Augusteijn et al.\ 1996; Beuermann et al.\ 1998).

Montgomery (2001) and Montgomery et al.\ (in preparation) developed
analytic expressions for both types of superhumps.  Montgomery et al.\
argued that in binaries having superhumps, the ratio between the negative
superhump deficit and the positive superhump excess
($\phi=\epsilon_{-}/\epsilon_{+}$) should be used to estimate the mass
ratio, rather than just one of these parameters.  Using their equation 9
and the observed $\phi$ parameter in TV~Col ($-0.357\pm0.055$), we can
estimate a mass ratio within the range 0.31-0.56.  The lower value is 
consistent with Murray's simulations.

% estimate for TV~Col

The mass ratio in TV~Col may thus be about the critical limit of 0.33.  A
criticism of this solution is that it suggests that it is only a
coincidence that TV~Col obeys the apparent relation between the superhump
period excess and the orbital period (Fig.~6).

\subsubsection{Invoking the magnetic field of the white dwarf}

Another solution to the problem might come from the suggestion that the
strong magnetic field of the white dwarf pushes the particles in the outer
accretion disc to larger orbits, allowing a disc size bigger than the
normal tidal radius in a non-magnetic system (this was confirmed by Murray,
personal communication).  In this way tidally unstable discs in systems
with mass ratios larger than 0.33 can be obtained.  We are, however, not
familiar with any model calculations that have been made for this case.

\section{Summary and conclusions}

\begin{enumerate}

\item We have found a 6.3-h period in the optical light curve of TV~Col in
the multi-longitude photometric campaign held in 2001 January.  This
detection confirms our findings from re-analysis of previously published
photometric data.  The periodicity is most naturally explained as a
positive superhump.

\item In the middle of the 2001 January run a short-lived minor outburst
occurred.  The new period is seen only in the data following outburst. The
outburst also changed the phase of the negative superhump by about a
quarter of a cycle, while its amplitude was slightly decreased.

\item Our findings support the classification of TV~Col as a permanent
superhump system.  Its superhump period is the longest known among CVs.
TV~Col thus offers a unique opportunity to test and reject some of the
models, as it extends the superhump regime to periods far beyond the
predicted values into a regime where the differences between models become
significant.  The mass ratio of TV~Col might exceed the limit for superhump
systems allowed by hydrodynamic simulations.

\item The observational and physical link between positive and negative
superhumps is thus strengthened by our result.  We might speculate that all
permanent superhump systems may have both types of superhumps.

% \item TV~Col offers a unique opportunity to test and reject some of the
% models, as it extends the superhump regime to periods far beyond the
% predicted values into a regime where the differences between models become
% significant.  The mass ratio of TV~Col might exceed the limit for superhump
% systems allowed by hydrodynamic simulations. Therefore, we have carried out
% spectroscopic observations in the infrared to determine reliably the binary
% masses and to confront the mass ratio with the predictions from the theory.
% We hope that these data, when analysed, will help solve this problem.

% We are excitingly looking forward to the reduction and analysis of these 
% data.

% \note{what spectroscopy?  This is the first mention of this!}

\item Our finding significantly extends the upper limit of orbital periods
in positive superhump systems (from 3.7 h to 5.5 h).  This range of periods
is populated with dwarf novae and nova-like systems.  Therefore, we
strongly urge observers to search for superhumps in nova-likes and dwarf
novae with orbital periods above the period gap and up to at least 5.5 h,
to check whether TV~Col is unique.  If TV~Col has certain properties that
can explain its extraordinarily long superhump period (e.g. classification
as an intermediate polar), then other long-period superhump systems should
possess similar features.

%3. A signal was also detected at the location of the X-ray spin period,
%however only in one of the best data sets.

\end{enumerate}

\section{Acknowledgments}

This paper uses observations made at the South African Astronomical 
Observatory (SAAO). We thank P. Woudt for his kind assistance as the 
support astronomer, J. Patterson for utilizing the CBA net and Michele 
Montgomery for sending us an early version of her paper prior to 
publication. Two anonymous referees are acknowledged for valuable 
comments. We also thank D. O'Donoghue for sending some of the data to 
CH and for using his reduction software. TN was a PPARC advanced fellow 
when the majority of this work was carried out. AR was supported by 
PPARC and is currently supported by the Australian Research Council.

% and AR were supported by 
% PPARC when the majority of this work was carried out. AR is currently 
% supported by the Australian Research Council.

% TN was a PPARC advanced fellow when the majority of this work was carried out.
% AR was supported by PPARC 
% during most of the period this work was carried out and is currently 
% supported by the Australian Research Council.

\end{document}